\documentclass{article}
\usepackage{amsmath}
\usepackage{amsfonts}
\usepackage{amssymb}
\usepackage{latexsym}
\usepackage[title]{appendix}
\usepackage{comment}

 \usepackage{eso-pic}

\title{Quantifying Resource Use in Computations
}

\author{R.J.J.H. van Son\footnote{
ACLC/University of Amsterdam, 
Spuistraat 210-212, 
1012 VT Amsterdam,
The Netherlands,
R.J.J.H.vanSon@gmail.com. Licensed under the Creative Commons Attribution license}
}

\begin{document}


\maketitle

\begin{abstract}
\noindent
It is currently not possible to quantify the resources
needed to perform a computation. As a consequence, it 
is not possible to reliably
evaluate the hardware resources needed for the application of 
algorithms or the running of programs. 
This is apparent in both computer science, for instance,
in cryptanalysis, and in neuroscience, for instance, comparative
neuro-anatomy. 
A System versus Environment game formalism is proposed based 
on \textit{Computability Logic} that allows to define
a computational work function that describes 
the theoretical and physical resources needed to perform any 
purely algorithmic computation. Within this formalism, the cost of a 
computation is
defined as the sum of information storage over the steps of the computation.
The size of the computational device, eg, the action table of a
Universal Turing Machine, the number of transistors in silicon, or the 
number and complexity of synapses in a neural net, is explicitly 
included in the computational cost. The proposed cost function 
leads in a natural way to known
computational trade-offs and can be used to estimate the computational
capacity of real silicon hardware and neural nets.
The theory is applied to a historical case of 56 bit DES key recovery, 
as an example of application to cryptanalysis. 
Furthermore, the relative computational capacities of human brain neurons 
and the C. elegans nervous system are estimated as an example of application
to neural nets.
\\
\noindent
keywords: computation, compatibility logic, neural nets, cryptanalysis
\end{abstract}

%
\section{Introduction}

In June 1998, a high ranking USA official, Robert S. Litt, testified before a
Senate judicial subcommittee that
\textit{...decrypting one single message that had been encrypted with a 56-bit
[Data Encryption Standard] key took 14,000 Pentium-level computers over four months;
obviously these kinds of resources are not available to the FBI}.
Later the same year, a 56 bit DES key was recovered in 56 hours at a cost of less than
\$250,000 using 1536 custom chips \cite{EFF_DES}.

The DES example points to the lack of a computational work function as
a fundamental problem in the theory of algorithms and
computation. At the time, questions were raised about the security
of 56 bit DES. In this debate, there was no way to estimate the resources needed 
to find a 56 bit key based on the available technology. So the above predictions
could neither be supported nor defeated in a quantitative way, except by going to the
expenses of actually cracking the keys.

A decade later, there still is no theoretical model for the abstract computational 
needs, or costs, of running an
algorithm, nor a way to evaluate the computational capacity of customized
hardware. This problem crops up more generally in game theory,
eg, when defining costs in computational Nash equilibria \cite{Halpern-2008,halpernPass-2008},
and in computational complexity theory when modeling
time and space bounded automata \cite{daylight-2008,Fortnow02ashort}. In a practical sense,
those who want to perform extensive computations have few tools to evaluate the 
computational power that current technology could (theoretically) provide.

At the other end of the spectrum of computational devices,
neuro-informatics studies how neural networks and the brain compute \cite{gorban-2003}.
There is an acute interest in understanding how nervous systems compute behavioral 
responses to environmental challenges \cite{gorban-2003,Connoretal2009,Lehrer2009,Eisenstein2009}.
Brain imaging and activity recording techniques, eg,
fMRI, MER, and ERPs, can show subsets of neurons computing
specific mental functions in real time. The local and long range connections between
neurons can be mapped in detail \cite{Connoretal2009,Lehrer2009,Eisenstein2009}. 
The underlying questions are \textit{what} is computed \textit{where}, and \textit{how}?
One obvious intermediate question is what can actually be
computed by a certain subset of neurons in a certain animal in a given time?
This is again a question on resource use in computations, but now
based on neurons instead of silicon gates.

In principle, it should be possible to compare the computational
capacities of the nerve systems of different animals like it is possible to compare
their metabolic rates. A human brain has on the order of $10^{11}$ neurons, whereas 
the nematode Caenorhabditis elegans has only 302 neurons in total 
(adult hermaphrodite, eg, \cite{CangelosiParisi1997}).
But how can the computational work these different neurons perform be compared? 
This is a question that is currently impossible to formulate in a quantitatively
meaningful manner.

The remainder of this paper is structured as follows. In section \ref{SectRequirements},
a model is proposed for quantifying the resource use, or cost function, for performing a 
computation  on theoretical  and real  devices (see also the Appendix). This model is 
applied to examples from cryptanalysis and neural 
physiology in section \ref{SectApplications}. The results are discussed in 
section \ref{SectDiscussion}. 

\section{A computational work function}\label{SectRequirements}

Any universal computational work function should have a few general features. 
It should describe
the resource needs of a computation in terms of costs. It should be abstract
enough to be applicable to both theoretical and real devices. It must
be able to add and remove resources during a computation. The cost must increase
strictly monotonically and must be additive in serial and parallel computations.
And finally, it must be possible to emulate any computational device efficiently,
where ``efficient'' is formalized here as a linear cost dependency. An efficient emulator allows
comparisons between different devices by comparing the sizes of emulator 
programs, independent of the emulated devices.

First, in section \ref{SectGameModel} a game model  of computation will be formulated 
that identifies resources and deliminates what is part of the computation for which 
the costs must be calculated and what is not.
Section \ref{SectCostFunction} proposes a cost function which has the desired 
features. The proposed cost function defines a least-cost implementation
for any computation for which an algorithm is known, which is explained in
section \ref{SectLeastCost}. The cost function is then used to model the computational
resources of
silicon hardware (section \ref{SectRealHardware}) and neural networks 
(section \ref{SectRealWetware}).

\subsection{The \textit{computability logic} game model}\label{SectGameModel}

Real computations need some material structure to carry and process the information, 
time and energy to allow state changes and to remove state information
while increasing the entropy of the environment
\cite{Levitin2007,SLloyd02_Capacity,SLloyd05_quant-ph0501135,SLloyd-2000_Limits}. 
So it is important to check whether the physics of computation does set 
limits on the resources needed in terms of the time, energy, and temperatures that are
required and entropy that is generated.
Of these factors, the minimum amount of energy $E$ to drive a bit sized state change in
time $\Delta t$ is $E \geq h/\Delta t$ and the minimum dissipation needed to
erase a bit is of the order $\Delta E \geq kT\ln(2)$, with $h$ Planck's
constant, $k$ the Boltzmann constant, and $T$ the absolute temperature. These
values are important on a molecular scale, or in quantum computers, but not in 
current computers \cite{SLloyd-2000_Limits}. So this study will ignore these physical 
constraints.

A theoretic framework that describes the qualitative
use of resources in computations well is \textit{computability logic}
\cite{japaridze2005,GiorgiJaparidze08012006}.
In computability logic, computability is defined in terms of games.
The ``computer'', or System, plays against the Environment
and ``wins'' if it can complete the requested computation successfully 
using the available resources.
This game model of computability explicitly defines what the responsibilities
of the Environment are and how it interfaces with the System. It also accounts 
for the resources that are used by the 
System to perform the computation and how the system communicates the results. 
Therefore, it is very well suited to delimit and define the costs
of computations.

Ignoring purely physical constraint, e.g., absolute time, temperature, and energy, 
in the cost function allows the 
use of a purely algorithmic game model from computability logic \cite{japaridze2005,GiorgiJaparidze08012006}. 
On this game model, a 
computational work function can be defined analogous to the
cryptanalysis work function of Shannon \cite{Shannon1949}.

The current study will restrict itself to such a purely algorithmic and deterministic
games where the speed of the moves is not relevant and
the environment has unlimited capacities to execute moves \cite{japaridze2005,GiorgiJaparidze08012006}. 
In the framework of 
computability logic, the System doing the computation is
further simplified by describing it as a collection of 
\textit{Universal Turing Machines} \cite{Turing1936CNA},
UTMs, each with a Finite State Machine, FSM, doing the processing and
three or more tapes: one or more work tapes, a valuation tape, and a run tape.

The work tape(s) correspond(s) to the working memory of a computer and
contains the program and all related data in use. The run tape corresponds to an
input/output medium that stores the moves written to it by the
System and the Environment. The System can not move backwards on the
run tape. That is, the System must use its own memory and cannot use the
(free) run tape to store the in- and output history.
The valuation tape contains the game specific parameters supplied by
the Environment and used by the program. A more general interpretation of
the valuation tape is that it contains any public information outside the
control of the System.

The System can recruit as many computational devices, UTMs, as it wants by specifying
them on the run tape. Every daughter device of the System can itself play against the
Environment on its personal run tape and receives a personal valuation tape.
Both the personal run and valuation tapes of each daughter device will be
copies of the original System tapes.
The communication between
the UTMs that make up the System is modelled by simply letting their work 
tapes overlap, but other solutions are possible.
Any UTM request should consist of a full
description of the finite state machine, initial state, contents of the work
tape, position of the heads, and the overlap between work tapes.

The computational model is completely interactive, so there are no general
rules limiting what can be written to the run tape. To make the resource use
explicit, it will be assumed that all moves are written as either fixed size or
self delimited strings.
Scanning the run tape for moves of the Environment is a computational
cost that must be born by the System. To minimize that cost, the moments
at which the Environment can write to a run tape are restricted.
The Environment will only write to a
run tape in response to a move of the device that ``plays'' on that run tape.
Any daughter device of the System will go to sleep
after it has written a move, and it wakes up only after the Environment has
responded. The computational costs are defined on the work tape(s) and the 
processing units (UTMs), but \textit{not} on the valuation and run tapes.

\subsection{A simple cost function}\label{SectCostFunction}

A very simple work, or cost, function for a single UTM that has all the above
features is
\begin{equation}\label{CostFunction}
C = \sum_{\lambda=1}^{\Lambda} I_{UTM}(\lambda)
\end{equation}
Where $C$ is the cost of a computation, $\Lambda$ is the number of steps needed
to complete the computation, and $I_{UTM}(\lambda)$ is the information in bits,
stored in that UTM at step $\lambda$ ($\lambda \leq \Lambda$).
In a situation with parallel UTMs, the cost is calculated for each UTM
separately using the step cycles of that UTM. Shared memory is attributed
to the UTM that makes the most steps. 

The cost function in equation \ref{CostFunction} replaces memory or time limited computations
with a limitation in $time \cdot memory$  (c.f., \cite{daylight-2008,Fortnow02ashort}).
$I_{UTM}$ is an information measure that is linear in its components
and always $I_{UTM} > 0$ for any computation in progress. Therefore, $C$ in equation 
\ref{CostFunction} is strictly monotonically increasing over ``time'' for any computation. 
The cost of a computation under equation \ref{CostFunction} is linear in time and 
computational resources. So the cost of doing computations in parallel on different 
computational devices or in series on a single device is simply the sum of the costs 
of doing the individual computations in isolation (provided the Environment takes 
care of initialization of the System between computations).
So equation \ref{CostFunction} has indeed the compositional features requested above.

The information $I_{UTM}(\lambda)$ is the information needed to specify
a UTM in the current state. That is, the information needed to specify at step
$\lambda$ the
\begin{itemize}
\item action table
\item the current state
\item the position of the heads
\item the current contents of the working tape
\end{itemize}
The working tape of a UTM is potentially of infinite size. But at any moment of
time, only a finite part of it is in actual use. For the cost calculations,
it is assumed that only part of the working tape is actually
``in use'' and the contribution of each work tape cell is proportional to
$\sim \log_2(N)$ (where $N$ is the total number of possible symbols). 
Memory locations are considered ``in use'', and part of the cost equation if
they have been written to during initialization or during operation of the
UTM. 

This can be compared to the System ``leasing'' new stretches of tape as needed. 
It is here assumed new memory automatically enters equation \ref{CostFunction} when 
an empty cell is written to. Some means for ending the ``lease'', i.e., ``freeing up'' 
tape is allowed.  This could simply be a special request on the run tape with an 
indicator of the working tape cells to be freed (eg, $X$ cells from the current head 
position).  After such a request, the specified part of the work tape is not part of the 
cost  equation anymore. The valuation and run tape
are not factored in, as these are considered part of the Environment.

In a game context, the output moves of a UTM  are only valid in a certain
context where, in some sense, the output symbols get a \textit{meaning}.
To be able to compare the costs of a computation using different UTMs, they must
all adhere to the same language on the output. A rigorous definition of the
cost of running a program can most easily be given on a single computational device.
An efficient emulater bridges the gap between different computational devices
For every finite set of UTMs, it is straightforward to define a UTM that can efficiently emulate
them all (see Appendix \ref{EfficientEmulator}). 

If the cost of doing a computation on the original UTM in $\Lambda$ steps
was $C$, then the cost of doing that computation on the emulator, $C^\prime$ will be:
\begin{equation}\label{EmulatorCost}
C^\prime \leq 4 \cdot \left( C  + \Lambda (\alpha + \epsilon) \right) + \beta
\end{equation}
The constants $\alpha$ and $ \beta$ are specific for the emulator whereas
$\epsilon$ is the ``rounding error'' of representing the original symbols and states
in the symbols of the emulator. All three constants can be determined from the
emulator program and structure. Examples of efficient emulators for UTMs and 
neural nets are given in Appendix \ref{EfficientEmulator}.

The cost function in equation \ref{CostFunction} incorporates several trade-off
relations. Most notably, a trade-off between processor complexity and length 
of computation in steps. A more complex computing device that
processes more bits per step can reduce the cost of a computation if the
memory use is large and vice versa. A specific case consists of a more
complex device that can reduce the number of steps in a computation without
increasing the amount of memory used. In such a case, the most efficient
set up would be to select a processor with a size that is
comparable to the average size of the memory used, 
$I_{device} \sim I_{\text{eff}}$ (see Appendix \ref{TradeOff}).

The cost function of equations \ref{CostFunction} emphasizes a drawback of standard UTMs. No 
practical computer will enumerate all memory positions to access a specific 
memory site, as a standard UTM
does, as this is not cost effective. Therefore, it will be assumed here that
the UTM can extract a relative address from the action table that will let the
head skip a number of cells on tape in a single clock cycle (i.e., processing step). This 
ability is related to the indirect addressing of register machines (e.g., Random Access 
Stored Program, RASP, or RAM  machines). 

Instead of adding a head skip with every entry in the action table, one or more 
accumulator/index registers could be added 
with some special states to manipulate them. However, the UTM with skip uses relative 
addressing, i.e., move head $i$ cells forward or backward, with a limited maximal skip. 
Furthermore, the relative position of the head over the tape is not explicitly stored (as a symbol)
and is not accessible to the System. It might depend on the computation and UTM
formulation whether the cost of the added complexity of the registers would be offset 
by the benefits.

The maximal number of cells that can be skipped in a single step affects the size 
of the action table, and the number
of states and symbols, so this ability does not come for free.
Going back from a UTM with $N$ symbols which allows for $D$ skipped cells 
to an equivalent UTM with only single cell moves, requires adding ``move'' states
which remember the original state and read symbol, and move one step. The addition
of these move states increases the total
number of states needed by a factor $O(N\cdot D)$ and computation time by a factor
$O(D)$. So the cost of a computation without skipped cells grows by a factor 
$O(N\cdot D^2)$ compared to a UTM with upto $D$ skipped cells (ignoring logarithmic terms). 

To make the cost of performing a computation on a UTM complete, the cost of 
operating the read/write head of the UTM should be taken into account. The structure
of the head follows directly from the action table. So the head of a UTM does not have to be 
specified separately. However, the head is the actual processing element and as
such constructing and operating one adds costs to a computation. A model of the 
computational cost of a UTM head is presented in Appendix \ref{HeadCost}.
The head is specified by the action table, and, for larger systems, the cost of operating 
the head is generally smaller than the costs associated with the action table. Therefore,
the contribution of the head to the cost of computations is ignored in the current study.

A quantitative example of complete cost calculations is presented in 
Appendix \ref{TitForTat} for a minimal Tit-for-Tat game.

\subsection{Least-cost implementation}\label{SectLeastCost}

A least-cost implementation can be defined in the same way as the algorithmic
or Kolmogorov complexity \cite{Chaitin69,LiVitanyi261084}. If a program $P$ is known that can
perform a computation on a UTM in finite time, then the least-cost program
can be found in a finite time too. The procedure is very simple and based
on the fact that a program with a size larger than $C$ cannot run for a single
step using less than $C$ resources. Run the original program $P$ and determine
it's cost $C$. Now run all programs $p_i$ with sizes smaller than $C$ 
(a finite number of programs) and stop
them if they have consumed $C$ in resources. All of these programs will stop
executing either because they halt on their own, or because they overrun the
cost limit. The program which needs the least resources to complete the 
computation is by definition the least-cost program.

To be able to compare different computational devices, eg, UTMs, all devices are
required to generate their output in the same alphabet. This
fits in the game formalism which requires the game participants to communicate
in a shared language of ``moves''. In the current context, a least-cost
combination of $\{P,UTM\}$ can be defined within the set of UTMs that can be
emulated by a specific emulator. The cost $C$ to be minimized
is that of equation \ref{EmulatorCost}. In this setting, a program on tape and
a ``program'' inside the processing unit become interchangeable.

The same procedure used
between programs on a single UTM can now be repeated over all UTMs. Any UTM with a
FSM size larger than $C$ cannot run even a single step within fixed cost bounds of $C$.
Determine the set of all UTMs with a size of their FSM $S \leq C$. This is a
finite set and can be emulated efficiently on a single device 
(see Appendix \ref{EfficientEmulator}).
Run each of them with all programs $p_i$ with a size $I_{p_i} + S < C$ until
they halt or have consumed $C$ in resources. Again, all these
programs will stop. Select the pair $\{P,UTM\}$ which consumed the least
resources as the least-cost option.

\subsection{Relations with real hardware}\label{SectRealHardware}

The cost function of equation \ref{CostFunction} is set in terms of stored
information times number of steps the information is used. For non-storage hardware,
this translates to
the information put into the device, in terms of components and connections, and
the operating frequency, ie, time per step.
That is, the hardware of the computational device is treated as a ``program''.
The connections between the active elements, eg, transistors, are
``programmable'' to the degree they can be freely chosen during design.

Although it might be difficult to model a modern complex CPU in terms of
component UTMs, it is possible to estimate the computational resources they
generate by looking at the transistor counts. As the cost function only looks at
memory ``use'', the CPU complexity can be reduced to the information needed to
describe the CPU state. That is, the variable state of the transistors and the
fixed structure of the connections, ie, $I_{CPU} \approx \log_2(\# states) +
\log_2(\#possible\ connections)$. It will be assumed, rather arbitrarily, that
transistors are mainly connected locally (small world topology) and each
transistor could on average have been connected in a hundred different ways
($\sim 7$ bits). Also, a transistor has a 1 bit state, \textit{on} or \textit{off}, 
and the size of the
``state machine'' is ignored as it can be covered by the state+connections.
Under these assumptions these numbers are $\log_2(\# states) =
O(\# transistors)$ and
$\log_2(\#possible\ connections) = O(\# transistors) \cdot 7$. Taken together,
each transistor is guessed to contribute around $\sim 8$ bits to $I_{CPU}$. This
is, of course, just a very crude, ball-park estimate. It is straightforward to
estimate the size of real computer systems from these principles.

As an example, the computational capacity of an off-the-shelf 2007 desktop system is estimated.
An AMD 64 X2 CPU core is made up of around 50 million transistors,
corresponding to $\sim 50\cdot 10^6$ byte of memory running at 3 GHz (2007,
source Wikipedia). So the resources produced by two such cores on a CPU could be
estimated at $\sim 3 \cdot 10^{17}$ byte/second. 2 GB high speed dynamic RAM
running at 400 MHz produces around $8 \cdot 10^{17}$ byte/s. It is rather
difficult to quantify magnetic disks, as it is not immediately clear what
clock-speed would be most appropriate.  A terabyte disk system would need a
$10^5$ Hz clock speed to get in the same order of magnitude as the other
subsystems, so it will be ignored for the moment. The on-chip caches are
small in comparison ($\sim 10^{15}$ byte/s) and will be ignored here too.
All together, a modern system with dual-core CPU and 2 GB RAM will run at around
$\sim 10^{18}$ byte/s, ie, at around 1 exabyte/s.

These data for general purpose CPUs can be compared to other types of devices.
Recently, GPUs (Graphical Processing Units), originally designed to render graphics
in personal computers and game consoles, are becoming popular in high performance 
computing \cite{StromGPU2009, Valich2009}.
A GPU can have half a billion transistors and runs at a half GHz with many parallel 
on-chip modules (data from 2007). For instance, the NVIDIA GeForce 8800 GT chip set 
contains 750 M transistors and runs at 0.6 GHz (source, Wikipedia, fall 2007). The 
crude metrics used here puts such a GPU at delivering $4.5 \cdot 10^{17}$ byte/s 
without memory. This is close to half what a AMD 64 could deliver, but optimized for 
its task.

According to these measures,
the original IBM PC with an Intel 8088 CPU (5 MHz, 29,000 transistors) and
0.1 MB memory would come in at about $\sim 10^{12}$ byte/s. Given the growth of
computing power, decibels would seem to be a more convenient measure of resource
size for a single computer in byte/second, eg, $10 \cdot
\log_{10}(I_{device}/10^{12})$, using the scale of the original IBM PC as a
reference. A dual-core AMD 64 system with 2GB RAM would then count as $\sim 60$
dB. Of course, equation \ref{CostFunction} cannot be expected to reflect cost
differences in real monetary terms. $60$ dB over 23 years (1984-2007)
corresponds to an increase of roughly $2.6$ dB/year.

\subsection{Relations with neurons}\label{SectRealWetware}

The same models as described above can in principle also be used to estimate the
capacity of neurons in the brain. However, in neurons it is not yet clear what anatomical
scale, and therefore, temporal scale, would be relevant to computation:
the cell, the synapse, or even the neurotransmitter receptor. In addition,
the current knowledge of neural computational functions and their relation to
the neuro-physiology is fragmentary at best. Therefore, the estimates described below
are only intended as illustrations of how the computational capacity
of real neural nets might be modelled.

Assume the synapse is the relevant active element \cite{ArcasFB03,ThomasGrant2009}
(``... a neuron is defined by synaptic connections'' \cite{ThomasGrant2009}).
Synapses are the contact points between neurons and it is generally
believed that they mediate most of the computational and learning activity of the nervous
system. The neuroanatomy of the human brain is far from settled \cite{Connoretal2009,Eisenstein2009,Lehrer2009} 
and it is difficult to put numbers on the populations of neurons and synapses with
any precision. For this example, only general estimates will be used as can be found
in textbooks. And the estimates will be limited to connections using chemical synapses.
Each neuron receives input from up to $10^4$ synapses (eg, \cite{MariebHoehn2007}). 
There are approximately $10^{11}$ neurons in the human brain (e.g., 
\cite{Lehrer2009}). So there are around $10^{15}$ synapses
in a human brain. In general, a synapse will originate from a local, nearby, neuron.
Take this local set to contain around $10^6$ neurons, which corresponds to 
connection distances of around 2 millimeters. The relative position
of a synapse on the neural body is important for its function. For simplicity,
the spatial structure
of the neuron is reduced to the relative position of the synapses. Both the
pre-synaptic and the post-synaptic part of the synapse can be in several (many)
states describing it's sensitivity to incoming action potentials and it's
ability to (de)polarize the post-synaptic membrane. As a last factor, the
runtime delay of incoming action potentials will differ between different
axon end points of the originating neuron. These differences have to be
modeled too.

The above description treats the synapse as a static, passive, device and 
the estimates are in line with \cite{DavelaarAbelman2006}. But 
biological neurons are dynamic, active, devices. This aspect of synaptic
function is important to computations \cite{PanticTK01}. This means that the
computational capacity should include the complexity of the synaptic ``device''.
At the moment, it is completely unclear how the size in bytes of the complexity
of the synapse should be estimated from physiological data. 

\section{Applications}\label{SectApplications}
\subsection{Understanding the DES cracker example}\label{SectEFFDESS}

The above theory might in future help support an informed discussion about
the potential capabilities of modern computer hardware. That way,
it might become less necessary to implement costly demonstrations
just to show that a certain prediction is wrong, like the one presented by the FBI 
analysts. 
Looking at the DES cracker 
example from the Introduction, it is possible to estimate the computational
resources available to the FBI and others at the time \cite{EFF_DES}. 

The protagonists in the example used two well known approaches to estimate the
costs of performing a computation. The public FBI approach was to take 
off-the-shelf systems, and estimate the run time and number of systems 
needed to perform the computation. The EFF approach was to design special 
purpose hardware and determine empirically what the requirements are in
terms of number of systems and run time. The current study tries to
base estimates on a combination of these approaches. This is done by
trying to estimate what performance could be achieved if the most complex
or powerful hardware available could be redesigned and optimized for
the desired computation. That is, first estimate what, according to 
equation \ref{CostFunction}, the maximum computational costs are that can 
be handled by existing hardware in a given time on any computation 
(the maximal performance). Next  estimate what the minimum cost is to perform 
the desired computation on  optimized hardware. Then compare these two under
the assumption that the existing hardware could be redesigned to be as
good as the optimized hardware.

A 1998 Pentium II
processor would have contained around $7.5 \cdot 10^6$ transistors and ran at
400 MHz. This would account for approximately $3.0 \cdot 10^{15}$ byte/s. A high
end system in 1998 would have up to 256 Mbyte of 100 MHz main memory, which
equates to  $2.6 \cdot 10^{16}$ byte/s. This brings the whole system up to
around $3 \cdot 10^{16}$ byte/s. 14,000
Pentium computers running for 4 months deliver $4.4 \cdot 10^{27}$ byte (steps).

The Electronic Frontier Foundation, EFF, succeeded in designing a search
unit in silicon that could check a 56 bit DES key in 16 clock cycles
\cite{EFF_DES}. The EFF were able to fit 24 such search
units onto a single chip with around 10,000 transistors and use the units in
parallel to check all possible keys. Many such chips can be used in
parallel. Using the earlier ball-park estimate of a contribution to $I_{CPU}$ of
8 bit per transistor, the computational effort
for a single encryption can therefore be estimated as $16 \cdot 10^4 / 24 \approx 6.7
\cdot 10^3$ byte (ignoring memory).

As one of
the design goals of DES was easy implementation, this low figure should not be a
surprise. If a general office computer of 1998
would have been a very efficient DES encryptor for its complexity, it would have
been able to test $4.5 \cdot 10^{11}$ keys a
second (again, ignoring memory). A single such computer should find a key in
less than 30 hours.

To evaluate the DES cracker, the housekeeping, communication and other functions
are deliberately ignored. Attention is focused on the key search. The DES
cracker chip could run with a clock
speed of 40 Mhz. In total, 1536 chips were used each with around 0.5 Mbyte of
memory. Together, this is $10^4 \cdot 4 \cdot 10^7 \cdot 1536$ or around $6.1
\cdot 10^{14}$ byte/s for the chips and $1536 \cdot 0.5$ Mbyte memory on 40 MHz or
only $3.1 \cdot 10^{10}$ byte/s for the memory. Together, these specialized
chips produce less as a computational resource than a single Pentium computer
of the time, or less than the
workstation used to coordinate the search. Running for 56 hours, the DES cracker
chips delivered $1.2 \cdot 10^{20}$ byte (steps). From this it can be concluded
that the DES cracker set-up was seven orders of magnitude more efficient in
DES encryption than a conventional computer of the day. Which is not really
remarkable given the simplicity of the DES encryption algorithm.

Obviously, general office
computers are all but efficient DES encryptors. Basically, the EFF used the fact
that silicon is equivalent to a program: it is
relatively easy to ``program'' a new chip to do exactly what is needed. If the
FBI analysts \cite{EFF_DES}, or their critics, had been
able to factor in the simplicity of the DES algorithm and the complexity of
hardware of the time, they would have been better able to
predict the vulnerability of the DES encryption.

\subsection{Comparing human and C. elegans neurons}\label{SectCelegans}

To describe each human brain synapse, an estimated 20 bits are needed 
to address the originating neuron out of a potential local 
population of 1 million.
Some 10-13 bits might be needed to indicate the synapse's relative 
position on the neural body. These 10-13 bits incorporate some of
the spatial organization of the neural body.
8 bits each are allocated for the pre- and post-synaptic states,
which might be a conservative estimate, given the complexity of synapses 
\cite{ThomasGrant2009}. The timing differences 
between synapses originating in
the same neuron could be described in, eg, 4 bits. So a conservative estimate
of the information needed to uniquely describe each synapse would be around 
50 bit, or in the order of 6 byte.
In total, on the order of $6\cdot 10^{15}$ byte (six petabyte)
are needed to describe the state
of all the synapses in a human brain. This is in accordance with the 
$\sim 10^{15}$ bit of \cite{DavelaarAbelman2006} (Note that the estimate in 
\cite{Wangetal2003} is unphysical as it exceeds the Beckenstein
bound for a brains sized object \cite{SLloyd-2000_Limits})

The number of neurons is four orders
of magnitude less than the number of synapses, so their contributions 
to the number of states are ignored. 
Action potentials have a maximum rate
of approximately 500 Hz. So it would be prudent to estimate the step
timing of synapses in the same range. That would mean that a conservative
estimation of the human brain indicates that it
calculates at a rate of $3 \cdot 10^{18}$ byte/s.

The above estimations are based on a static synapse model. In reality, synapses
are dynamic entities that adapt to stimulation \cite{PanticTK01}. 
It is estimated here that two bytes are needed to describe the state of the 
synapse. To simplify matters,
it is assumed that 10 bits of these are needed to describe dynamic state 
parameters. To get at least an order of magnitude estimate, the action table size 
of a UTM with the same number of states, $2^{10}$, is used as a proxy measure. 
From this it follows
that the complexity of the synapse is of the order of $10^3$ bytes (on the order of 
$\sim$10 bits per state). This increases the
estimated capacity of the human brain to something in the order of $10^{21}$ byte/s.

Compare the human central nervous system to the neural system of C. elegans \cite{CangelosiParisi1997}. 
An adult hermaphrodite contains
302 neurons and around 7000 synapses. Each neuron has on average around 25 incoming
synapses. That is, the originating neuron can be described in 8 bit and the position of the 
incoming synapse on the neural body in around 4 bits. Timing differences in incoming synapses 
can probably be ignored (0 bit). It is unclear how the pre- and postsynaptic state information 
relates between nematodes and mammals, but here it is arbitrarily assumed that nematodes will 
need less bits, just to put a number on it, 10 instead of 16 bits. In total, around 22 bit 
would be needed to completely describe the state and position in each synapse in a nematode, 
or less than 3 bytes. This is assuming only static synapses. Again, the contribution of the 
neurons is partly included in the post-synaptic state, and partly ignored.

As nematodes are not homeiotherm, the switching speed of the synapses will be lower than in 
mammals. For a ten degrees difference in body temperature ($37^\circ$ versus $25^\circ$ C), 
at least a halving of the metabolic rate, and switching speed, is expected. Using only the
values for static synapses, the nervous system of a complete hermaphrodite adult C. elegans 
would then have a computational capacity of 
$7000 \cdot 3 \cdot 0.25 \cdot 10^3 \sim 5\cdot 10^6$ byte per second. A single human 
neuron would have $10^4$ synapses each needing around 6 bytes to describe statically, 
working at $0.5 \cdot 10^3$ Hz for a total of $3\cdot 10^7$ byte per second. So, according 
to these crude, ballpark, estimations, a single human brain neuron processes, or computes, 
around six times as much information than the complete neural system of a C. elegans adult.

It is informative to look at what makes individual human neutrons perform at a higher level 
than the complete neural system of C. elegance. The important factors are 1) number of
synapses, 2) population of possible originating neurons, 3) spatial interactions between
synapses on a neuron, 4) metabolic speed.

1) The number of synapses ending on a single human neuron and in the complete
C. elegance body are comparable (7,000 versus 10,000). As it is assumed that the 
synapses are the computational entities, this 
fact alone predicts comparable performance. 

2) Each synapse in C. elegance can originate in some 300 other neurons. This corresponds
to some 8 bit to describe the possible information processing wirings. Each human synapse
can originate, potentially, from $10^{11}$ other neurons. Here it is assumed that connections
in the human brain are in general local (a small world network) and the real, or effective, 
number of originating neurons in the human brain is much more limited. But any 
realistic number for the human brain  will be way larger than the 300 in C. elegance. 
In our example this is simply limited to a million originating neurons, i.e., 20 bits. But even 
with only a 20,000 possible originating neurons this would still be double the contribution
of a C. elegance synapse. 

3) With $\sim 10,000$ synapses contacting each human neuron
compared to the 25 synapses contacting each C. elegance neuron,
the options for spatial interactions between synapses increases. In our simple
model this increases the computational power of human neurons from 
approximately 4 to 13 bits. 

4) Last, there is an expected metabolic speed doubling, 
from 25$^\circ$C to 37$^\circ$C, which would double computational performance.

\section{Discussion and conclusions}\label{SectDiscussion}

Almost from the start of the computer era, questions about the time and memory needed
to complete a computation were raised \cite{Fortnow02ashort}. A lot of theoretical
progress has been made towards these questions in the fields of game theory, logic, and 
computational complexity. The current study tries to bring these developments a step closer
to the practical developments in other fields, eg, cryptanalysis, neuro-imaging,
and neuro-informatics. A pressing need in these latter fields is an
evaluation of the computational resources of an actual processor, eg,
the electronic hardware or neuronal wet-ware, and to link these to the
theoretical powers of Turing Complete theoretical devices, eg, the UTM.
This inclusion of the processing hardware in the accounting of the resources
is a challenge which requires a way to valuate \textit{memory} and 
\textit{processing hardware} in a uniform currency that can be integrated 
with, or in, \textit{time}.

Based on a few natural requirements, a simple formula for 
a computational work function for quantified
resource use emerges with the features of \textit{Memory} times
\textit{Steps}, ie, a dimension of bytes (equation \ref{CostFunction}). 
In more intuitive physical terms, the computational resources
are counted as an integration of \textit{information} (\textit{entropy})
over a \textit{normalized interaction time}. This count includes the information
frozen into the computational device itself, eg, the UTM action table, the
silicon of the CPU, or the neurons and synapses in a nervous system.

This definition of the cost of a computation directly leads to the concept of a least-cost
implementation, both for a single computational device and
between devices. Such a least-cost implementation can always
be found within a finite time given a single example program that can perform the
computation. As such, the least-cost is a universal invariant of the computation.

In the end, computing is done using some physical substrate. This substrate,
eg, silicon chips or neural tissue, will need to have some, non-random, 
structure to be able to run a program, eg, transistors, 
synapses, and most of all connections. The information stored in this 
structure, as far as it is relevant to computations, is the $I_{device}$
needed to calculate the computational costs of equation \ref{CostFunction}.

Reducing silicon CPU complexity to concrete hardware design features
like transistor count and connectivity, and memory capacity, it is possible to 
roughly guess both the capacity of real computer hardware and the 
hardware needs of (simple) algorithms. A more refined model, 
that takes into account what level of complexity can
be achieved by custom hardware, can be used to estimate the real costs of
implementing and executing abstract mathematical algorithms. 
This would allow, for instance,
security analysts to be better prepared to the increasing power of computer
hardware than they currently are.

The same models can also be used to estimate the
capacity of neurons in the human or animal nervous system. 
These estimations are currently rather speculative.
But it can be easily shown on elementary neuro-anatomical
and neuro-physiological arguments that each individual human 
brain neuron should outperform the complete nervous system 
of C. elegans by almost an order of magnitude. 

It is even possible to compare the computational capacity of neural nets and silicon. 
However, this does not lead to a lot of insight immediately. 
Neurons and silicon
are on different ends of the computational spectrum.
The computational capacity of silicon is dominated by it's clock speed.
On the other hand, neurons are slow, but the capacity of neural nets
is dominated by their connectivity. This formalizes the well known 
fact that the computational strengths of human brains and 
silicon computers lie in
completely different problem areas. Simulating the one in the other
has always proved to be extremely inefficient.

\section{Funding}
Netherlands Organization for Scientific Research 
(276-75-002)

\bibliographystyle{hep}
\bibliography{CryptAttacks}

\newpage

\begin{appendices}
\renewcommand{\theequation}{\thesection-\arabic{equation}}

\section{Efficient emulators}\label{EfficientEmulator}
\setcounter{equation}{0}  

The Efficient Emulator requirement can be defined as follows:
Given a finite set of computational devices of a certain 
complexity, eg, UTMs
with up to $t$ tapes and action
table sizes of unto $S$ bits, that can perform computation
$A$ in $\Lambda$ steps with cost $C$, there exists a computing device which can
emulate any of these devices performing $A$ at a cost
$C^\prime$ such that (by definition)
\begin{equation}\label{EfficientEmulation}
C^\prime \leq \gamma \cdot \left( C  + O(\Lambda) \right) + O(1)\;\;\text{with}\
\gamma > 0
\end{equation}
For a UTM emulator, $O(\Lambda)$ can be interpreted as $\Lambda\cdot(\alpha + \epsilon)$ 
and $O(1)$ as $\beta$. For a UTM, $\alpha$, $\beta$, and $\gamma$ are fixed emulator cost factors
for all emulated devices and computations and $\epsilon$ represents the ``rounding error'' in
representing the original states and symbols on the emulated device, eg, UTM$_a$, in the 
symbols of the emulator, eg, UTM$_b$. The size of the rounding error $\epsilon$ can be estimated
from the encoding of the emulated device, eg, the action table.

Efficient emulation according to equation \ref{EfficientEmulation} is possible
using equation \ref{CostFunction} as a cost function at least for some Turing 
Complete devices. Which means that any algorithm
that can be computed efficiently by one device, eg, a UTM, can also be computed
efficiently by other devices. It is a weak condition as it does not ensure
that there will always be an efficient emulator of a specific type.

If the lowest cost, $C^\prime$, of a certain computation on any efficient
emulator is known, it can be shown that the cheapest program on any emulated device,
eg, UTM$_a$, that can
perform the same computation in $\Lambda$ steps has a cost $C$ of at least
\begin{equation}
C +  \Lambda\cdot (\alpha + \epsilon) \geq (C^\prime - \beta)/\gamma
\end{equation}
Where $C$ itself depends on $\Lambda$ (see equation \ref{CostFunction}).

Below, two examples of efficient emulators are given. One emulates all single tape UTMs up to a 
given number of symbols and states. The other emulates simplified neural nets unto a maximal
number of nodes and connections.

\subsection{An efficient emulator of UTMs}

For any UTM$_a$ with $\leq t$ tapes, it is possible to design
a $t+1$ tape UTM$_b$ that can emulate it efficiently.
Here this is proven for $t=1$, but
other cases and types of devices follow directly from this case.
The dual tape UTM$_b$ uses one tape, $T_a$ with head 
$H_a$, to store the
tape of UTM$_a$. The other tape, $T_b$ with head 
$H_b$, contains the action table of
UTM$_a$, organized as a table of \\
\{New symbol  $T_a$,  Move $H_a$, Move $H_b$\} 
addressed by row addresses 
\{State UTM$_a$, Symbol $\sigma_a$ on $T_a$\}.
The action table of UTM$_b$ has a simple structure, and will not be described here. 

At the
start of an emulated read-write-move cycle of UTM$_a$, the position of the
head, $H_b$, over the $T_b$ tape indicates the current state of UTM$_a$.
UTM$_b$ performs the following steps:
\begin{enumerate}
 \item Move $H_b$ to correct row on $T_b$ in stored action table of UTM$_a$
 \begin{itemize}
 \item Read current symbol $\sigma_a$ from tape $T_a$
 \item Move $H_b$ by $\sigma_a$ rows
 \end{itemize}
 \item Read and write new symbol
 \begin{itemize}
 \item Read new symbol $\sigma^\prime_a$ from $T_b$
 \item Write $\sigma^\prime_a$ to tape $T_a$
 \item Move $H_b$ to next field in row
 \end{itemize}
 \item Move $H_a$
 \begin{itemize}
 \item Read $H_a$ movement $D_a$ from $T_b$
 \item Move $H_a$ by $D_a$
 \item Move $H_b$ to next field in row
 \end{itemize}
 \item Move $H_b$ to new state of UTM$_a$
 \begin{itemize}
 \item Read $H_b$ movement $D_b$ from $T_b$
 \item Move $H_b$ by $D_b$
       to the row position that indicates the new state of UTM$_a$
 \end{itemize}
\end{enumerate}
The $Halt$ state of UTM$_a$ will move UTM$_b$ to a program tape area
that will halt UTM$_b$.

It is obvious that UTM$_b$ can only efficiently emulate single tape UTMs for
which it can handle all symbols, states, and head movements inside it's own tape
symbols. This sets an upper size limit to the UTMs it can emulate. But within these
size limits, this emulator clearly works according to equation
\ref{EfficientEmulation}. UTM$_b$ can emulate every single
read-write-move cycle of any UTM$_a$ in four of its own read-write-move
cycles ($\gamma = 4$). The action table of UTM$_a$ can be stored in
$N_a\cdot M_a$ rows of 3 symbols of UTM$_b$, which takes more
space than $log_2(N_a M_a D_a)$ bits by $\epsilon = O(1)$. The action table,
state and other work tape contents of UTM$_b$ are fixed, contributing
$\gamma \cdot \alpha (=O(1))$ per emulated read-write-move cycle of UTM$_a$ for a
total of $\gamma \cdot \Lambda \alpha$ ($= O(\Lambda)$). 
Starting and halting costs are also of order $\beta = O(1)$.

\subsection{An efficient emulator for neural nets}\label{EfficientNeuron}

An efficient emulator for a simplified neural network can be build from parallel processors.
Such an emulator will be generated as above for a UTM. In a general game
model, a UTM is recruited for each neural node and
then a UTM for each synapse or connection between neural nodes. There are a maximum
of $N_{max}$ nodes available each with at most $k$ incoming connections (synapses).
In total, a maximum of $k\cdot N_{max}$ synapses will be available. Unused nodes 
and synapses are unconnected and have empty worktapes, but they do contribute to the
computational cost of the emulator.
All UTMs have dual work tapes,
$T_\alpha$ stores the node and synapse states and $T_\beta$ a state transformation
table. The $T_\alpha$ work tapes of the
synapse UTMs will overlap with a shared state field in the $T_\alpha$
work tape of the  \textit{originating} neural node UTM and a "personal" 
activity field in the
\textit{target} neural node UTM.
It is assumed that an unlimited number of UTMs can read concurrently from
the same field of a shared work tape. However, only one UTM at a time can
write to a shared field.

Each synapse UTM has a table that tells how a current synapse state $\sigma$ changes
to a new state $\sigma^\prime$ under influence of the state, $\eta$, of the originating
node ($0$ or $1$, for firing a spike or not). The table contains a
row for every possible synapse state. Each row contains fields: \\
\{Activation, Head movement for $\eta = 0$, Head movement for $\eta = 1$\} \\
The relation between the synapse state (row number) and the Activation
is the weight of the synapse. ``Learning'' could result in changing the
activation entries (not implemented here).

Start at the work tape position of the head $H_\alpha$ over $T_\alpha$
that contains the state of the originating node and head $H_\beta$ over the
start of the table row on
$T_\beta$ that contains the current state of the synapse. Then first update all the
synapse UTMs in parallel.
\begin{enumerate}
\item Synapse UTMs read the state of the originating node UTMs
      \begin{itemize}
      \item Read node state $\eta$, which is either $0$ or $1$ (fire spike)
      \item Move $H_\alpha$ to next field containing the current activation
      \item Move $H_\beta$ to the row field corresponding to the node state $\eta$
      \end{itemize}
\item Read new synapse states $\sigma^\prime$ from $T_\beta$
      \begin{itemize}
      \item Read $\sigma^\prime$ from $T_\beta$ as a relative head movement $D_\beta$
      \item Move $H_\beta$ by $D_\beta$
      \end{itemize}
\item Read and write activation of corresponding synapse states
      \begin{itemize}
      \item Read Activation from $T_\beta$
      \item Write Activation to $T_\alpha$
      \item Move $H_\alpha$ to previous field
      \end{itemize}
\end{enumerate}
Then update the neural node UTMs. The head, $H_\alpha$ starts at the 
first Activation field
(of $k$ fields) on $T_\alpha$.
$T_\beta$ contains a table to relate the new state to the activation level.
The table is organized in rows with \{New State, New Activation\}. The new activation
level which follows the current, is stored as a movement of $H_\beta$.
The position of $H_\beta$ indicates the current activation of the node.
\begin{enumerate}
\setcounter{enumi}{3}
\item \label{Integration} Sum activation fields ($k$ steps), end over node state field
      \begin{itemize}
      \item Read activation $\rho$ from $T_\alpha$
      \item Move $H_\beta$  by $\rho$ rows
      \item Move $H_\alpha$  to next field
      \end{itemize}
\item \label{NeuralNonlin} Read new state and update node state
      \begin{itemize}
      \item Read new state $\eta^\prime$ from $T_\beta$
      \item Write new state $\eta^\prime$ to $T_\alpha$
      \item Move $H_\beta$ to next field on $T_\beta$
      \end{itemize}
\item Update node activation state
      \begin{itemize}
      \item Read new activation state from $T_\beta$
      \item Move $H_\alpha$  back to first activation field (ie, by $k$ fields)
      \item Move $H_\beta$  to new activations state $\eta$\\
            (eg, start of the current row for state $0$, and to the start
            row of the table after state $1$, spike generation)
      \end{itemize}
\end{enumerate}
With the exception of step \ref{Integration}, all steps take a single
cycle of the UTMs. In total, a single cycle on the original neuron can be 
emulated in $\gamma = k+5$ steps of all the UTMs in parallel. 

Step \ref{Integration} is extremely inefficient because it needs
$k$ steps to sum the activations, and every neural network 
solves the problem by using a fast \textit{integrator}. Such an integrator will
sum the synapse activations in a short time. This integration can
be done by a fast or parallel accumulator. 

Assume that all activation symbols
are two's complement bit numbers (to allow for inhibitting synapses) that
indicate the size of the activation. The accumulator would contain a register
with $A$ bits representing the current activation and an adder with $A$ full 
bit adders. A one bit full adder has a truth table of $8 \cdot 2 = 16$ bits, 3 bits for
inputs and carry-in to indicate the row and 2 bits for output and carry-out.
The first and last bit adders need only half as much, 8 bits, because they lack a
carry-in or carry-out.

If the truth tables are used as the complexity of the adders, the $A$ bit
accumulator would need $3 \cdot A - 1$ bit registers (accumulator, input, and 
$A - 1$ carry bits) and $16 \cdot (A - 1)$ bit truth tables, or, $19 \cdot A - 17$ 
bits. In this calculation the two half bit adders are combined. 

So for a 32 bit activation size, the accumulator
would need around 591 bits. A parallel integrator can be simulated by a
fast accumulator which sums the $k$ activation fields in a single clock step.
That is, step \ref{Integration} is performed in a single step by the
accumulator which sums all the activations and prints out a selection of bits
from the accumulator (not necessarily all $A$ bits). The cost of such a fast 
accumulator would be $k \cdot (A \cdot 19 - 17) = k \cdot I_A$ per clock step. 
Note that this is approximately the same cost as would be needed for $k$ parallel
accumulators working in a single step. Then step \ref{NeuralNonlin} is changed
to read the activation and generate the spike (1) or not (0). 

The original cost, $C$, of a computation of
a neural network with $N$ nodes and $k$ synapses per node, from $N$ 
originating nodes, over $\Lambda$ steps is
\begin{equation}
C =  \Lambda N \left( I_{node} + k (I_{syn} + \log_2(N)) \right)
\end{equation}
Where $I_{node}$ and $I_{syn}$ are the total information size of the
neuron nodes and synapses, respectively. Here, the complexities are
estimated as the sizes of the action tables of equivalent UTMs, as there
is currently no sensible estimate based on physiological data.
With a fast accumulator, the simulation of a node splits the complexity
into an accumulator part ($k \cdot I_A$) and ``the rest'' ($I_{B}$), ie,
$I_{node} = I_{B} + k \cdot I_A$.

To calculate the cost of the emulation, using the fast accumulator, the sizes of the
emulator UTMs and accumulator without tapes are $\alpha_{B}$, $\alpha_A$, and 
$\alpha_{syn}$ for emulating the node body, accumulator, and synapse, respectively.
The corresponding rounding errors for emulating the real neural states and symbols 
in the emulator are $\epsilon_{B}$, $\epsilon_A$, and $\epsilon_{syn}$.
For $N$ nodes with each $k$ synapses and $\log_2(N)$ bits to designate the
originating node, the emulator cost becomes:
\begin{eqnarray} \label{EmulatedNeuron}
C^\prime &\leq&   6 \Lambda N (I_{B} + \epsilon_{B} +
                        k  (I_{syn}+\log_2(N)+ I_A + \epsilon_{syn} + \epsilon_A)) +\beta \\
                       & &  +\ 6 \Lambda N_{max} (\alpha_{B} + k_{max} (\alpha_A + \alpha_{syn})) \nonumber \\
         &\leq&   6 C + 6 \Lambda N (\epsilon_{B} + k  (\epsilon_A + \epsilon_{syn})) 
                           + 6 \Lambda N_{max}(\alpha_B + k_{max} (\alpha_A + \alpha_{syn})) +\beta \nonumber
\end{eqnarray}
The cost in equation \ref{EmulatedNeuron} is indeed linear in $C$, and $\Lambda$ 
according to equation \ref{EfficientEmulation}, for fixed maximum $N_{max}$ and $k_{max}$.

\section{Complexity versus time trade-off}\label{TradeOff}
\setcounter{equation}{0}  

Using a more complex Finite-State-Machine (FSM) often reduces the time and 
cost needed to complete a lengthy computation. On the other hand, moving a 
short computation to a smaller device can reduce costs too.
The boundaries of such trade-offs follow from the cost function.
As an example, consider a UTM$_\alpha$ with $M$ states, $N$ symbols and $D$ possible
head movements. UTM$_\alpha$ has a FSM size $S = MN(m + n + d) + m$, where $m, n, d$
are the bit sizes needed to store, respectively, states, symbols, and head movements.
Assume there is an efficient, low-cost, program $P$
for UTM$_\alpha$ that computes $A$ in $\Lambda$ steps effectively using 
$I_{\text{eff}}$ bits on
tape (where $\Lambda I_{\text{eff}} \equiv \sum^\Lambda_1 I(\lambda)$) with cost
$C\simeq\Lambda (S + I_{\text{eff}} + \log_2(I_{\text{eff}}))$. At each step, 
UTM$_\alpha$ can process $b=m + n + d$ bits.
Assume that $A$ depends on the total number of bits processed, $\Lambda \cdot b$.

Construct a new UTM$_\beta$ that can process $b^\prime$ bits per step, or
\begin{equation}
b^\prime = \delta b = \delta_1 m + \delta_2 n + \delta_3 d
\;\; \text{with}\;\; \delta > 0
\end{equation}
and take $\{S,I_{\text{eff}} \}\gg \{b, \log_2(I_{\text{eff}})\}$. 
Very simple examples of such operations would be to combine
program steps for parallel execution to increase $b$, or to split 
program steps into smaller components to decrease $b$.

The new UTM$_\beta$ is chosen such as to reduce the
cost of computation $A$.
UTM$_\beta$ has a FSM size, $S^\prime$, of
\begin{equation}
S^\prime = M^{\delta_1}N^{\delta_2}\delta(m + n + d) + \delta_1 m \approx \delta M^{\delta_1 - 1}N^{\delta_2 - 1} S
\end{equation}
Simplify the new FSM size to $S^\prime = \delta \Gamma_{\delta} S$ where
$\Gamma_{\delta} \equiv M^{\delta_1 - 1}N^{\delta_2 - 1}$ can be roughly 
approximated as an exponential function of $\delta$,
$\Gamma_{\delta} \sim 2^{(m+n)(\delta - 1)}$.
A new, efficient, program $P^\prime$ on UTM$_\beta$ can calculate $A$ by
processing $\Lambda^\prime \cdot b^\prime \geq \Lambda \cdot b$ bits or in
$\frac{1}{\delta}\Lambda \leq \Lambda^\prime \leq \Lambda$ steps. As the total number of
bits processed remain the same, it is assumed that $I_{\text{eff}}^\prime \geq I_{\text{eff}}$.

The cost, $C^\prime$, of computing $A$ using $P^\prime$ on UTM$_\beta$ can be estimated as
$C^\prime \geq \Lambda^\prime (S^\prime + I_{\text{eff}}^\prime + log_2(I_{\text{eff}}^\prime))$. After ignoring
small components $\delta_1 m$ and $log_2(I_{\text{eff}})$, the new cost becomes
\begin{eqnarray}\label{EqCmplxCost}
C^\prime &\geq& \dfrac{1}{\delta}\cdot
		\dfrac{ \delta \cdot \Gamma_{\delta} S + I_{\text{eff}}}
		{S + I_{\text{eff}}} \cdot C
\end{eqnarray}
For large $I_{\text{eff}} \gg \delta \Gamma_\delta S$, the new cost becomes 
$C^\prime \geq \frac{C}{\delta}$ which is a decrease if $\delta > 1$. For
small $I_{\text{eff}} \ll \delta \Gamma_\delta S$, the new cost becomes 
$C^\prime \geq \Gamma_{\delta} C$ which is a decrease if $\delta < 1$. 
Note that in
the limits of $I_{\text{eff}} \rightarrow \infty$ and 
$I_{\text{eff}} \rightarrow 0$ the costs can be made very small indeed by, 
respectively, increasing or decreasing $S$.

The optimal size of the FSM can be estimated by calculating the minimum 
of equation \ref{EqCmplxCost}. Express the effective memory size in terms
of the FSM size, $I_{\text{eff}} = \omega S$ and assume that 
$\delta \approx \delta_1  \approx \delta_2  \approx \delta_3$. 
Differentiate with respect
to $\delta$. The minimum cost is reached if:
\begin{equation} \label{EqMinDelta}
 \omega = \delta^2 \cdot 2^{(m+n)(\delta - 1)} \cdot \dfrac{(m+n)}{\ln(2)}
\end{equation}
The optimal size of a FSM is reached if $\delta = 1$, which means that
the minimal cost is reached if 
$S =  \frac{\ln(2)}{(m+n)}\cdot I_{\text{eff}} $.

The above boundaries on the cost are for the ideal cases, where both the memory use, $I^\prime_{\text{eff}}$, 
as the number
of steps, $\Lambda^\prime$, are minimal. In general, a cost reduction to $1/\delta_{\text{eff}}$ can be
found for large, $I_{\text{eff}}$, if
\begin{equation}
\dfrac{\Lambda^\prime I^\prime_{\text{eff}}}{ \Lambda I_{\text{eff}}} \equiv \dfrac{1}{\delta_{\text{eff}}} < 1
\end{equation}
These results suggest that the optimum results are found for choices of $b$ for which $S  \approx I_{\text{eff}}/(m+n)$. This implies that $MN \sim O(I_{\text{eff}})$ for $d \sim O(m,n)$.

The above modelling refers to computations that are processor bound, ie, the computations depend 
on the number of bits processed. For such a computation, the most efficient
implementation should try to reduce the number of computational steps by equalizing the
complexity (``size'') of the central processor and the amount of memory used.

\section{The cost of operating the UTM head}\label{HeadCost}
\setcounter{equation}{0}  

In a UTM, the head is the ``processing element''. The head reads and writes symbols,
and steps forward and backward. It can also be seen as responsible for changing the 
state of a UTM.  The structure of the head
is fully determined by the actions table, ie, number of states, symbols, and 
possible head movements.
So it does not have to be specified in the definition of a UTM.
However, the complexity of the moving head adds to the real costs of
operating a UTM. 

The complexity of the UTM head can be estimated, in symbolic terms, 
from the number of symbols and head movements. For $N$ symbols, at least 
$n\ge\log_2(N)$ bits are needed for each of the read and the write functions. Head 
movements over the tape and state changes in the action table will be 
implemented as counters that keep track of the relative movements
over the tape and the action table and signals when zero is reached (count down). 

For each bit in a counter, 2 bits are needed for the register and carry-in,
2 bits for the output and carry-out and $4\cdot 2=8$ bits for the truth table.
In a counter, only the carry-in bits are counted as the carry-out bits are the same
bits.
In total, 11 bits are needed per counter bit. The last counter bit does not need a carry-out
bit and only needs a 4 bit truth table. So a counter  of width $w$,
needs $11w - 4$ bit of "content" for the bare counter. 
A compare-to-zero can be implemented as a logical OR over the $w$ bits of the 
counter that is triggered by the result 0/\textit{false}. This can be implemented by
an OR of each bit with the result of the higher order bits. For each bit, except
the highest order bit, two inputs and one output and a 4 
bit (OR) truth table are needed, where all but the last output are shared with the
next input. 
Together, 6 bits per counter bit for a total of $6(w-1)+1$ bits for a counter of width
$w$.  So a counter plus zero comparator with $w$ bits needs 
$I_{counter} = 17w - 9$ bits of logic storage. Note that the information needed
to describe the connections is ignored here for simplicity.

With $M$ states, the state counter into the rows of the action table needs a width of 
$m\ge\log_2(M)$ bits and a total  content of $17m - 9$ bit. To address the columns
in the action table with $N$ symbols, the counter width is $n \ge \log_2(N)$ with a
total content of $17n-9$ bit.
For a maximal range of $D$ steps, the tape counter will need $d \ge \log_2(D)$
bits width and a total content of $17 d- 9$ bits. With one bit dedicated to the direction
 of movement,
the latter might be reduced by 18 bit. For the purpose of generality, the full $17d-9$ 
bits will be used here. Operating a state or tape counter running $M$, $N$, and $D$ 
steps would cost, respectively, $M(17m-9)$,  $N(17n-9)$, and $D(17d-9)$ bit 
(steps) of our work function.

In total, $\sim 2 n + 17 (m+n+d) - 27$ bits are needed
to specify the state of the head during operation for a cost of:
\begin{equation}\label{eqHeadCost}
\dfrac{C_{head}}{ \Lambda}  = 2n+M(17m-9)+N(17n-9)+D(17d-9)
\end{equation}
Where $C_{head}$ is the cost of running the head in bits and $\Lambda$
is the length of the computation in clock steps. Equation \ref{eqHeadCost}
only represents the minimum cost in symbolic (bit) terms. 
 
Reducing the UTM head to one that does not
skip tape cells, $D =2$ (i.e., $\{-1,1\}$), increases the number of steps needed
to complete the computation by a factor of $O(D/2)$ and increases the number
of states needed, and the size of the action table, by a factor of $O(ND/2)$ to store
state and symbol information while stepping to the desired tape cell. 
So the cost of running the computation 
increases by a factor of $O(ND^2/4)$, both when accounting for the action table 
size and when accounting for the state counter cost
(ignoring logarithmic terms).
The tape  counter will run at approximate the same cost as the decrease in the number
of counts compensates for the increased duration of the computation, ignoring 
logarithmic factors. The  cost of the symbol read and  write heads and the symbol 
counter will increase by a factor $O(D/2)$ due to the longer compute times.

The cost of using a UTM can be divided into the size of the tape and action table, and the 
cost of deploying the head. For a fully 8 bit UTM, $m=n=d=8$, the head will account for
just over 6\% of the non-tape cost (107 kb versus 1.57  Mb for the action table), down to
under 0.03\% for a fully 16 bit UTM ($m=n=d=16$).

From the definitions it can be derived that, for large $N$ and $M$, the cost of
running the head becomes small compared to the cost of the action table if
$17(N+M+D) \ll NM$. This is satisfied if $D \le \max(M,N)$ and $\min(M,N) \gg 51$.
Both conditions are not unreasonable for practical systems doing long computations.
See Appendix \ref{TradeOff} for trade-offs between $M$, $N$, $I_{\text{eff}}$, and $\Lambda$.

The information to prescribe the UTM head can be extracted from the action table 
and does not have to be specified independently. Moreover, for UTMs which are not
minimalist, the contribution of the head to the costs of the computation will be 
relatively small.
To simplify this study, the contributions of the head to the costs of computations 
will, therefore, be ignored in this paper.

\section{A quantitative cost example: Tit-for-Tat}\label{TitForTat}
\setcounter{equation}{0}  

To illustrate the cost computations in the game model, it will be applied to the Iterated Prisoner's 
Dilemma game with Tit-for-Tat as the strategy \cite{RAxelrod03271981}.
The game is played on a single Run tape, where the moves of the System and Environment
are written in pairs of cells. 

There are three symbols: $C$ for cooperate, $D$ for defect
and $H$ for halt. The environment starts a turn by writing a string of two cells,
one with a random symbol $C$ or $D$, and one with the Environment's move, either $C$ 
or $D$.  Then the Environment wakes up the System which is always positioned on the first 
cell of the string where it is supposed to write a move. The System completes the turn in a two 
step cycle. Note that the System has \textit{no} private tape and cannot move backward over 
the Run tape. 

First, the system reads the content of the cell it is positioned over and writes its current move, 
$C$ or $D$, into the cell. If the symbol read was $H$, the System halts and the game is over, 
else the System moves the head to the next cell. In the next step, the System 
reads the symbol in the underlying cell, moves to the next (empty) cell and goes to sleep
(if that cell is empty). Then the Environment generates the next turn.

The Tit-for-Tat strategy is implemented in a simplified Turing Machine with five states: 
Cooperate ($c$), Defect ($d$), Read ($r$), and Halt ($h$). There are three symbols, 
$C$ (cooperate), $D$ (defect), and $H$ (halt). The System cycles through the turns as 
follows:
\begin{enumerate}
\item cycle
\begin{itemize}
 \item Wake up by Environment
 \item Read content of Run tape
\item Depending on the current state write:
\begin{itemize}
	\item[$C$] if state is $c$
	\item[$D$] if state is $d$
\end{itemize}
\item Move to next cell
\item If read symbol was:
\begin{itemize}
	\item[$H$] switch to $h$ $\rightarrow\ halt$ 
	\item[$C$] switch to $r$
	\item[$D$] switch to $r$
\end{itemize}
\end{itemize}
\item cycle
\begin{itemize}
 \item Read content of Run tape
 \item Do not write (or write back the read symbol)
 \item Move to next cell
\item If read symbol was:
\begin{itemize}
	\item[$H$] switch to $h$ $\rightarrow\ halt$ 
	\item[$C$] switch to $c$ $\rightarrow\ sleep$
	\item[$D$] switch to $d$ $\rightarrow\ sleep$
\end{itemize}
\end{itemize}
\end{enumerate}
The game starts by writing the specification of the Tit-for-Tat program on the
valuation tape with $c$ as the initial state. The Environment loads the program,
writes its first move and positions the System over the first cell on the Run tape
in state $c$, and starts the System. The game ends when the Environment
writes an $H$ symbol which halts the System.

The System goes to ``sleep'' when it reaches an uninitialized tape cell. If the 
Environment writes all its moves in one go, the System will not go to sleep
and play until it reaches an $H$ symbol. Else it will sleep until the cell under
its head is initialized.

The action table of the Tit-for-Tat player is  presented in table \ref{tableT4T}.
\begin{table}[b]
\caption{Action table for Tit-for-Tat player. The System is not actually allowed to
overwrite the move of the Environment. For completeness, the system is set to 
rewrite the Environment's move in state $r$.}
\begin{center}
\begin{tabular}{c|ccc|ccc|ccc}
          & \multicolumn{9}{|c}{Symbol read}   \\
State & $C$  &&&  $D$  &&&  H &&  \\
\hline
c        &$C$&1& $r$ & $C$ & 1&  $r$ & $C$ & 0 & $h$  \\
d        &$D$&1& $r$ & $D$ & 1&  $r$ & $D$ & 0 & $h$  \\
r        &$C$&1& $c$ & $D$ & 1&  $d$ & C   & 0 & $h$  \\
\hline
          & \multicolumn{9}{|c}{symbol write - move - new state}   \\
\end{tabular}
\end{center}
\label{tableT4T}
\end{table}
This simplified implementation is very small, 4 states and 3 symbols. After entering
the $h$ state the System halts and the game is over.  The System cannot write the 
$H$ symbol. There is a rule that the Environment
is not allowed to write an $H$ symbol in its second, move, cell because it would 
lead to an incomplete game. If the Environment does make this illegal move, it loses. 
The fact that the System writes a $C$ symbol in the same cell afterward (lower right 
hand side cell of table \ref{tableT4T}), effectively breaking the rule that is not allowed 
to change the move 
of the Environment, does not change this outcome. In \textit{compatibility logic}, 
the player who makes the \textit{first} illegal move loses. 

The total action table could fit in 36 bit ($3\cdot 4$ bits per row 
and 3 rows) and the current state in 2 bits. One 
turn would take two clock cycles.  The cost for the System 
of running the game would be 76 bits per turn ($2\cdot 38$) as sleep time is not counted. 
It is easy to see how the complexity of the System's game 
playing strategy can be increased by including one or more private work tapes and more states.
However, such a more complex strategy would increase the costs of the computation, potentially
by a very large amount.

If the cost of running the head is included, with $N=3$ symbols, $M=4$ states, and
$D=2$ movement options, the complexity of the run tape head would be
$n_w=1$ bit for writing (2 symbols) and $n_r\sim 2$ bit for reading (3 symbols), 
$17\cdot 1 - 9 = 8$ bit for the head movement counter, and $17\cdot 2 - 9=25$ bit 
each for the state change and symbol counter. 
The cost of running the head would be $2\cdot 219=438$ bit per move 
(per step: $n_w+n_r+M(17m-9)+N(17n-9)+D(17d-9) = 
1+2+4\cdot 25+4\cdot 25+2\cdot 8=219$ bit). So running the head would be 
the major cost of running this Tit-for-Tat Machine.

Some aspects of computability logic have been used implicitly in this example. 
Most notably the fact that any player who breaks the rules loses. So if any of the 
players would rewrite any of the moves, it would lose. The Environment can read 
any cell, and therefore, has to write its moves first in every turn or else it could cheat. 
The System cannot move back over the Run tape, so it has to write down its own 
move before it can read the Environment's move or else forgo this turn. It is a free 
design choice to go for a game structure that prevents this type of cheating instead 
of a rule to bind the players. Both approaches would work. It is the Environment 
who determines whether the System has completed the computation and, therefor, 
``wins''. This means the System is not required to keep track of the score, which 
would be costly. In this Tit-for-Tat game, the condition for winning could be anything 
from not breaking the rules to actually getting the most points.

\end{appendices}

\end{document}